\pdfoutput=1
\documentclass[11pt]{article}

\usepackage[preprint]{acl}

\usepackage{times}
\usepackage{latexsym}
\usepackage[T1]{fontenc}
\usepackage[utf8]{inputenc}
\usepackage{microtype}
\usepackage{inconsolata}
\usepackage{graphicx}
\usepackage{amsmath}
\usepackage{amssymb}
\usepackage{booktabs}
\usepackage{multirow}

\definecolor{bestblue}{RGB}{0,70,180}
\definecolor{secondred}{RGB}{190,30,30}
\newcommand{\best}[1]{\textcolor{bestblue}{\textbf{#1}}}
\newcommand{\second}[1]{\textcolor{secondred}{\textbf{#1}}}

\title{SheetCraft: Fine-Tuning Code Agents for Execution-Guided Spreadsheet Reasoning and Construction}

\author{
  \textbf{Zuowei Zhang\textsuperscript{1,*,$\dagger$}},
  \textbf{Xiaotian Li\textsuperscript{2,*}},
  \textbf{Siyuan Li\textsuperscript{3}},
  \textbf{Enhao Gu\textsuperscript{4}},
\\
  \textbf{Kexu Cheng\textsuperscript{5}},
  \textbf{Hanbo Song\textsuperscript{6}},
  \textbf{Yule Zhang\textsuperscript{1}}
\\
\\
  \textsuperscript{1}University of Science and Technology of China,
  \textsuperscript{2}Zhejiang University,
\\
  \textsuperscript{3}Harbin Institute of Technology, Shenzhen,
  \textsuperscript{4}East China Normal University,
\\
  \textsuperscript{5}Institute of Computing Technology, Chinese Academy of Sciences,
\\
  \textsuperscript{6}Beijing University of Posts and Telecommunications
\\
  \small{
    \textsuperscript{*}Equal contribution. \quad \textsuperscript{$\dagger$}Corresponding author: \href{mailto:zuowei.zhang@mail.ustc.edu.cn}{zuowei.zhang@mail.ustc.edu.cn}
  }
}

\begin{document}
\maketitle

\begin{abstract}
Spreadsheets are a common interface for operational data management, but building realistic spreadsheet artifacts requires coordinated edits across tables, fields, formulas, views, dashboards, workflows, and permissions. We study this problem as an execution-guided code-agent task. Given a natural-language request and an initial spreadsheet base, the agent must generate executable actions that transform the base into a persistent target artifact. This setting is different from table question answering and one-shot code generation because correctness depends on state transitions, cross-component consistency, and repair from concrete execution failures. We present \textbf{SheetCraft}, a training and evaluation framework for spreadsheet code agents that combines compilation feedback, sandbox execution, and artifact-level judging. SheetCraft trains on distilled tool-use QA data and multi-turn code-agent trajectories derived from real usage scenarios, allowing the model to learn both operation semantics and feedback-conditioned repair. On a 595-case held-out benchmark drawn from real usage scenarios with 1,750 user turns, SheetCraft-27B achieves 96.6/95.0 case-level all-turns compile/execute success and 98.9/98.3 turn-level compile/execute success. Further analysis shows that trajectory supervision, teacher diversity, and budgeted repair are key to reliable spreadsheet construction.
\end{abstract}

\section{Introduction}
Spreadsheets remain a primary interface for organizing operational data. Users often need artifacts that combine tables, typed fields, formulas, views, dashboards, workflows, and permission settings. Natural language interfaces can make this process easier, but the output is not a textual answer. As shown in Figure 1, the agent must transform an initial workbook into a persistent spreadsheet state that satisfies the user request.

\begin{figure*}[t]
    \centering
    \includegraphics[width=\textwidth]{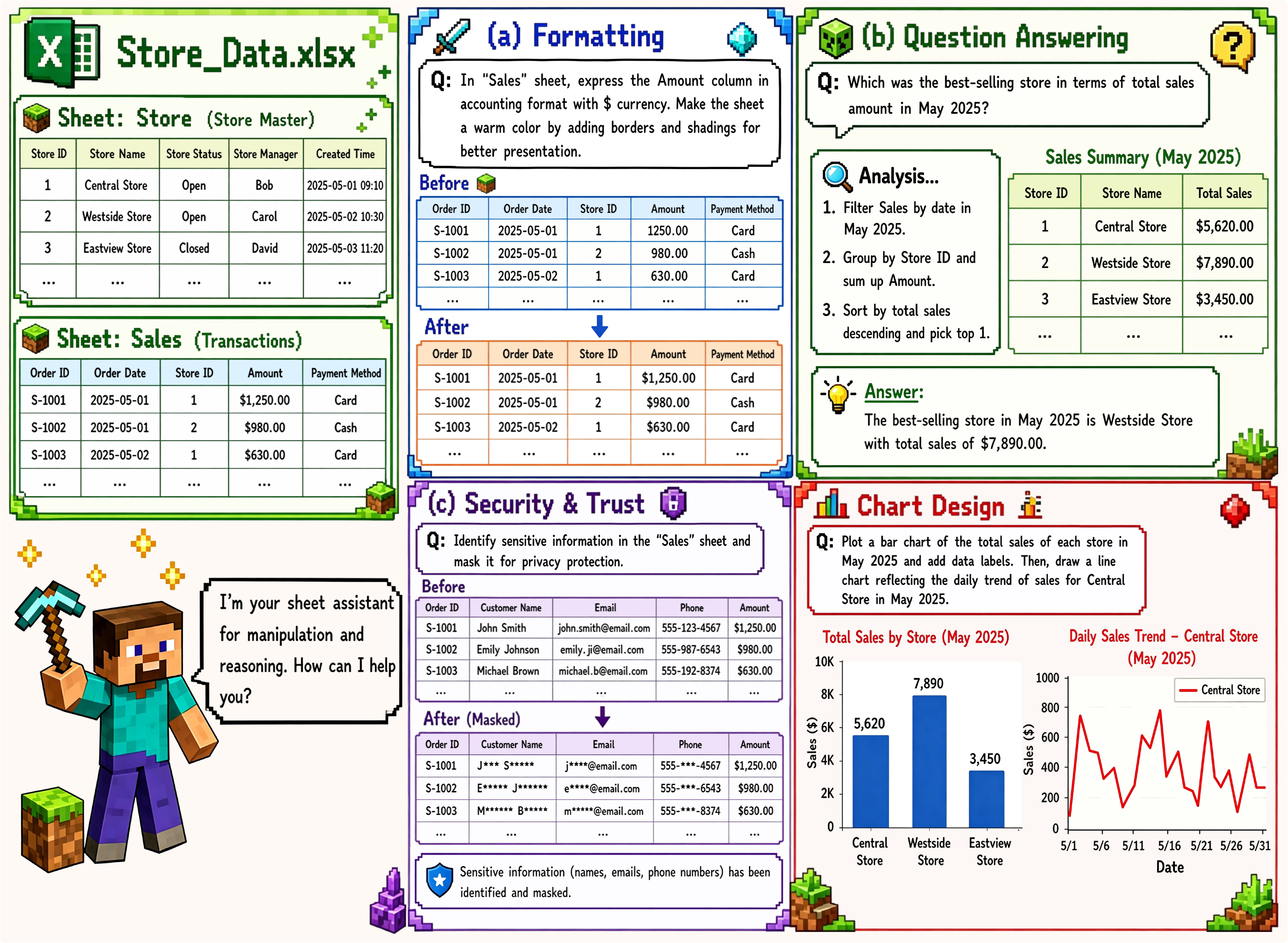}
    \caption{SheetCraft assists users with spreadsheet manipulation and reasoning tasks.}
    \label{fig:overview}
\end{figure*}

This task differs from conventional table understanding and question answering, where prior work often focuses on representing tabular inputs or returning denotations from static tables \citep{yin2020tabert,herzig2020tapas}. Spreadsheet construction is stateful and operational. The agent must identify relevant sheets and columns, infer underspecified intent, choose valid operations, and maintain consistency across dependent components. Small mistakes in schema design, field references, execution order, or formatting logic can invalidate the final artifact even when the generated code looks plausible.

We therefore formulate spreadsheet construction as an execution-guided code-agent task. Given a sequence of user requirements, an initial spreadsheet base, and a typed operation interface, the agent generates TypeScript programs that modify the base. Each attempt is checked by compilation and sandbox execution, and failures are converted into feedback for the next repair step. This process makes the model accountable to the artifact it produces, rather than to surface-level code fluency alone.

To study this task, we introduce SheetCraft, a framework for fine-tuning and evaluating spreadsheet code agents. SheetCraft uses staged supervision. Distilled QA examples teach operation-level tool knowledge, while multi-turn trajectories teach the model to repair compile and execution failures. At evaluation time, SheetCraft measures both intermediate compile/execute success and the final spreadsheet artifact, making it possible to test whether the agent actually constructs the requested state.

In summary, our contributions are as follows:
\begin{itemize}
    \item We formulate spreadsheet construction as an execution-guided code-agent task that maps natural language requirements into persistent spreadsheet artifacts.
    \item We develop SheetCraft, a budgeted multi-turn framework that combines supervised trajectory learning, compilation feedback, sandbox execution, and artifact-level judging.
    \item We construct distilled QA and multi-turn trajectory data from real usage scenarios and apply accuracy and privacy filtering before training.
    \item We evaluate SheetCraft on a 595-case held-out benchmark from real usage scenarios and analyze compile/execute success, Pass@$K$, domain-level artifact judging, and ablations.
\end{itemize}

\section{Related Works}

\paragraph{Tool-using language agents.} Recent work studies large language models as agents that reason, call tools, and react to observations. ReAct interleaves reasoning and actions \citep{yao2023react}; Toolformer studies self-supervised API use \citep{schick2023toolformer}; and API-Bank and ToolLLM provide benchmarks and instruction-tuning pipelines for tool-augmented LLMs \citep{li2023apibank,qin2024toolllm}, while NaviAgent introduces graph-driven bilevel planning for scalable multi-tool orchestration \citep{jiang2026naviagent}. More recently, trajectory-guided tool learning frameworks such as SpatialCLI further demonstrate how tool-use trajectories can be converted into supervision to internalize structured tool reasoning \citep{zhou2026spatialcli}. WebShop and SWE-agent further show that task-specific action spaces and environment feedback are important for interactive tasks \citep{yao2022webshop,yang2024sweagent}. These works motivate tool-using agents, but they mainly target API selection, retrieval, web interaction, or software editing rather than stateful spreadsheet construction. More broadly, retrieval-augmented grounding has also been explored to enforce domain-specific constraints in generative tasks \citep{cheng2026physrag}.

\paragraph{Feedback-guided code generation and repair.} Feedback is also central to program generation and repair. Program-aided language models use interpreters for reasoning \citep{gao2023pal}, execution-based verifiers rerank language-to-code outputs \citep{ni2023lever}, and AlphaCode shows the value of sampling and behavioral filtering for code generation \citep{li2022alphacode}. Reflexion and self-debugging use verbal or execution feedback for iterative improvement \citep{shinn2023reflexion,chen2024selfdebug}, while SWE-bench emphasizes realistic long-context software repair \citep{jimenez2024swebench}, and SWE-Touch further studies coding agents under dynamic workspace state changes when users concurrently edit the code \citep{tan2026swetouch}. Recent work on reinforcement learning with verifiable rewards also highlights the importance of controlling context and rollout budgets under verifiable feedback \citep{yangbatched}. Similarly, expert trajectory guidance has been shown to prevent superficially coherent yet inaccurate multi-step reasoning in progressive decision-making domains \citep{zhu2026medeyes}. Our setting adds a different constraint: the generated code must compile, execute against a cloned spreadsheet base, and leave behind a correct structured artifact. SheetCraft therefore grounds repair in spreadsheet-specific trajectories, diagnostics, sandbox execution, and artifact-level judging.

\paragraph{Table and spreadsheet reasoning.} Structured-data reasoning has moved from static table QA toward richer table manipulation. TaBERT and TaPas pretrain joint representations of language and tables \citep{yin2020tabert,herzig2020tapas}, while spreadsheet program-synthesis work automates transformations from examples \citep{gulwani2011automating}. Chain-of-Table modifies tables during reasoning \citep{wang2024chainoftable}, and TableLlama trains generalist models for table tasks \citep{zhang2024tablellama}. SpreadsheetBench and SheetAgent are closest to our setting because they evaluate spreadsheet manipulation and long-horizon spreadsheet agents \citep{ma2024spreadsheetbench,chen2025sheetagent}. SheetCraft focuses on supervised training of spreadsheet code agents from real-scenario QA data and multi-turn execution trajectories, and evaluates the constructed spreadsheet state through compile, execute, and artifact-level gates.

\begin{figure*}[t]
    \centering
    \includegraphics[width=\textwidth]{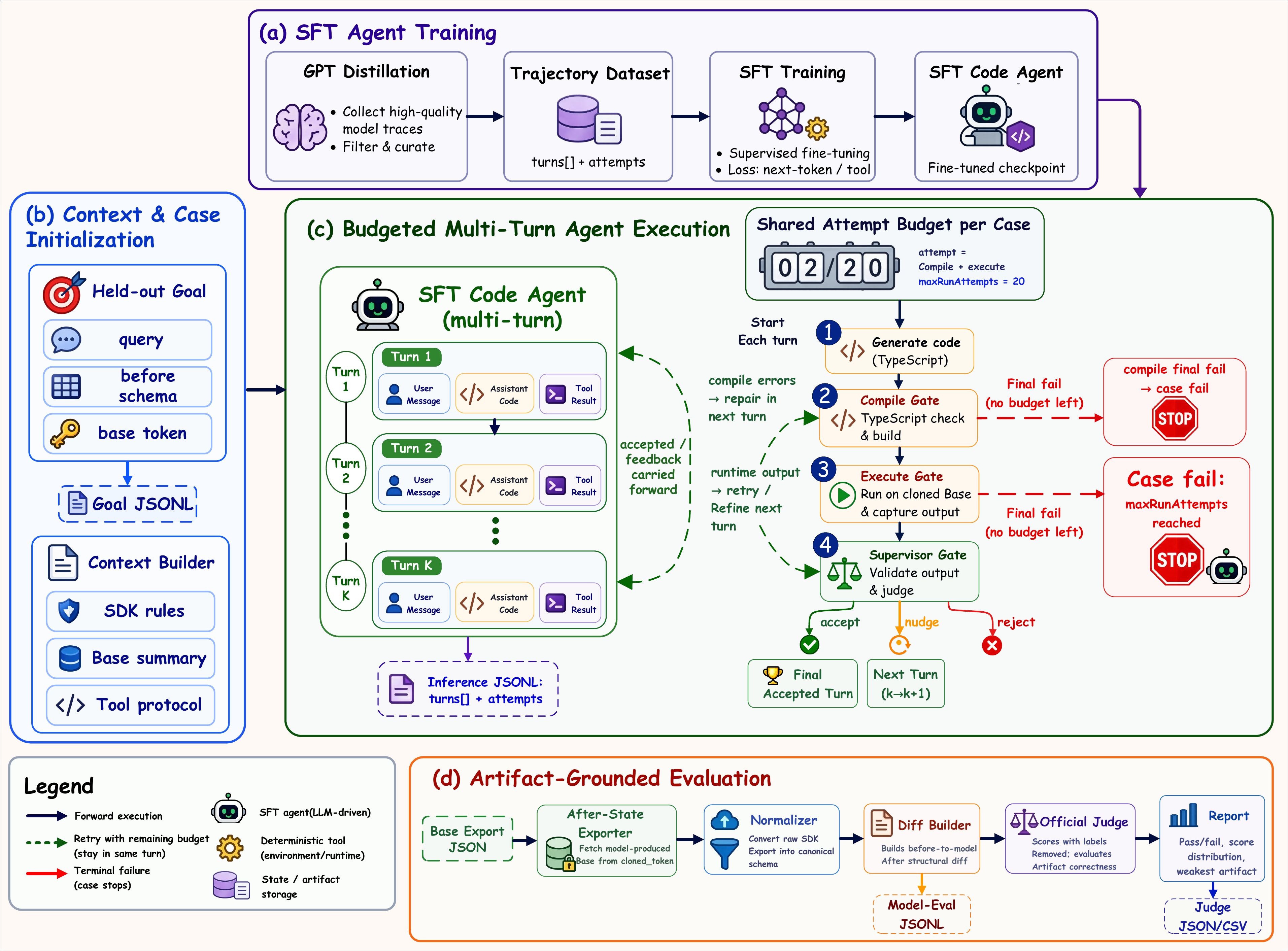}
    \caption{Overview of the SheetCraft training, budgeted multi-turn execution, and artifact-grounded evaluation pipeline.}
    \label{fig:pipeline}
\end{figure*}

\section{Method}

\subsection{Task Formulation and Overview}
Figure~\ref{fig:pipeline} summarizes the SheetCraft pipeline. We define a case as $c = (\mathbf{q}, B_0, \mathcal{T}, y)$, where $\mathbf{q} = (q_1, \dots, q_n)$ is a sequence of user requirements, $B_0$ is the initial spreadsheet base, $\mathcal{T}$ is the TypeScript operation interface, and $y$ is a hidden reference artifact. At turn $i$, the agent observes $h_i = (q_{\le i}, B_{i-1}, f_{<i})$ and generates a program $a_i$. When execution succeeds, the spreadsheet state is updated as
\[
B_i = \mathrm{Exec}(a_i, B_{i-1}), \quad i = 1, \dots, n.
\]
The goal is to construct a final artifact $B_n$ that satisfies both operational validity and user intent:
\[
\max_{a_{1:n}} J_{\mathrm{art}}(N(B_n), y), \qquad \prod_{i=1}^{n} C_i E_i = 1,
\]
where $C_i = C(a_i)$ and $E_i = E(a_i, B_{i-1})$ denote compile and execution success, $N(\cdot)$ canonicalizes the spreadsheet state, and $J_{\mathrm{art}}$ is the artifact-level judge. The task is therefore artifact-centric. A valid prediction must compile, execute, and create the requested tables, views, dashboards, workflows, permissions, and cross-component configurations.

\subsection{Data Distillation and Privacy Filtering}
We construct two data streams from real user spreadsheet-construction scenarios. Following instruction-tuning and synthetic-instruction bootstrapping practices \citep{ouyang2022instructgpt,wang2023selfinstruct}, $D_{\mathrm{qa}}$ contains operation-level QA examples distilled from GPT-5.5-Pro and GLM-5.1, while $D_{\mathrm{traj}}$ contains multi-turn code-agent trajectories from the same teachers. Each trajectory records the user request, spreadsheet context, operation documentation, generated TypeScript, compile or execution feedback, and final spreadsheet state.

Before SFT, we apply an accuracy filter $A(\cdot)$ and a privacy filter $P(\cdot)$:
\[
D_{\mathrm{sft}} = \{P(z) \mid z \in D_{\mathrm{qa}} \cup D_{\mathrm{traj}}, \; A(z) = 1\}.
\]

The accuracy filter removes malformed schemas, incomplete turns, invalid operation calls, inconsistent states, and programs that fail the required compile or execution checks. For trajectories, we keep repair attempts only when the feedback and the subsequent edit form a coherent correction path.

Because the data come from real usage scenarios, privacy filtering is required before training and packaging. This step is motivated by evidence that neural language models can memorize and expose rare or sensitive strings \citep{carlini2019secret,carlini2021extracting}. We run OpenAI Privacy Filter\footnote{\href{https://openai.com/index/introducing-openai-privacy-filter/}{OpenAI Privacy Filter official page}} over user requests, schemas, table contents, tool logs, and code comments. Detected names, emails, phone numbers, private links, tokens, and other identifiers are removed or replaced with typed placeholders, preserving task semantics while excluding user-specific sensitive content.

\subsection{Supervised Agent Training}
The upper part of Figure~\ref{fig:pipeline} shows supervised agent training. We train the model in two stages. The QA stage teaches operation semantics, signatures, argument constraints, and dependencies between spreadsheet components:
\[
\mathcal{L}_{\mathrm{qa}}(\theta) = - \sum_{(x, r) \in D_{\mathrm{qa}}} \log p_\theta(r \mid x).
\]
The trajectory stage teaches feedback-conditioned construction. A trajectory is $\tau = \{(m_t, a_t, o_t)\}_{t=1}^T$, where $m_t$ is the model context, $a_t$ is the assistant program, and $o_t$ is compile, execution, or alignment feedback. We optimize next-token likelihood over assistant program tokens:
\[
\mathcal{L}_{\mathrm{traj}}(\theta) = - \sum_{\tau \in D_{\mathrm{traj}}} \sum_{t=1}^{T} \sum_{u \in a_t} \log p_\theta(u \mid m_t, u_{<}).
\]
Keeping both successful programs and useful repair attempts teaches the model to preserve valid edits, react to diagnostics, and improve the spreadsheet state under a fixed attempt budget.

\subsection{Budgeted Multi-Turn Execution}
The middle part of Figure~\ref{fig:pipeline} shows inference. For each turn $i$, the system clones the current base, constructs $h_i$, and allows at most $K$ repair attempts. At attempt $k$, the model samples
\[
a_i^k \sim p_\theta(\cdot \mid h_i, f_i^{<k}).
\]
The program first passes a compile gate $C(a_i^k)$. If compilation succeeds, the program is executed in a sandbox to obtain $E(a_i^k, B_{i-1})$ and a candidate state $\widehat{B}_i^k$. Alignment checks then compare the program and the candidate state against the user request. The accepted attempt is the first $k$ satisfying
\[
C_i^k = 1, \quad E_i^k = 1, \quad R_i^k = 1,
\]
where $C_i^k = C(a_i^k)$, $E_i^k = E(a_i^k, B_{i-1})$, and $R_i^k = R(q_i, a_i^k, \widehat{B}_i^k)$ denotes lightweight intent and product checks. Failed compile, execution, or alignment attempts produce feedback $f_i^k$ that is appended to the next prompt. The fixed budget encourages targeted repair rather than repeated regeneration.

\subsection{Artifact-Grounded Evaluation}
The lower part of Figure~\ref{fig:pipeline} illustrates evaluation. We use a requirement judge and an artifact judge. The requirement judge scores whether the generated implementation satisfies the user request across domains such as tables, dashboards, workflows, and permissions:
\[
s_{\mathrm{req}} = J_{\mathrm{req}}(\mathbf{q}, a_{1:n}).
\]
The artifact judge evaluates the executed result. After termination, the final base is exported and normalized into $N(B_n)$. Normalization removes incidental ordering, transient identifiers, and interface metadata while preserving task-relevant structure. The final score is
\[
s_{\mathrm{art}} = J_{\mathrm{art}}(N(B_n), y).
\]
Because $y$ is hidden during inference, this protocol tests whether the program constructs the intended spreadsheet artifact, not merely whether the code appears plausible.

\section{Experiments}

\begin{figure}[t]
    \centering
    \includegraphics[width=\columnwidth]{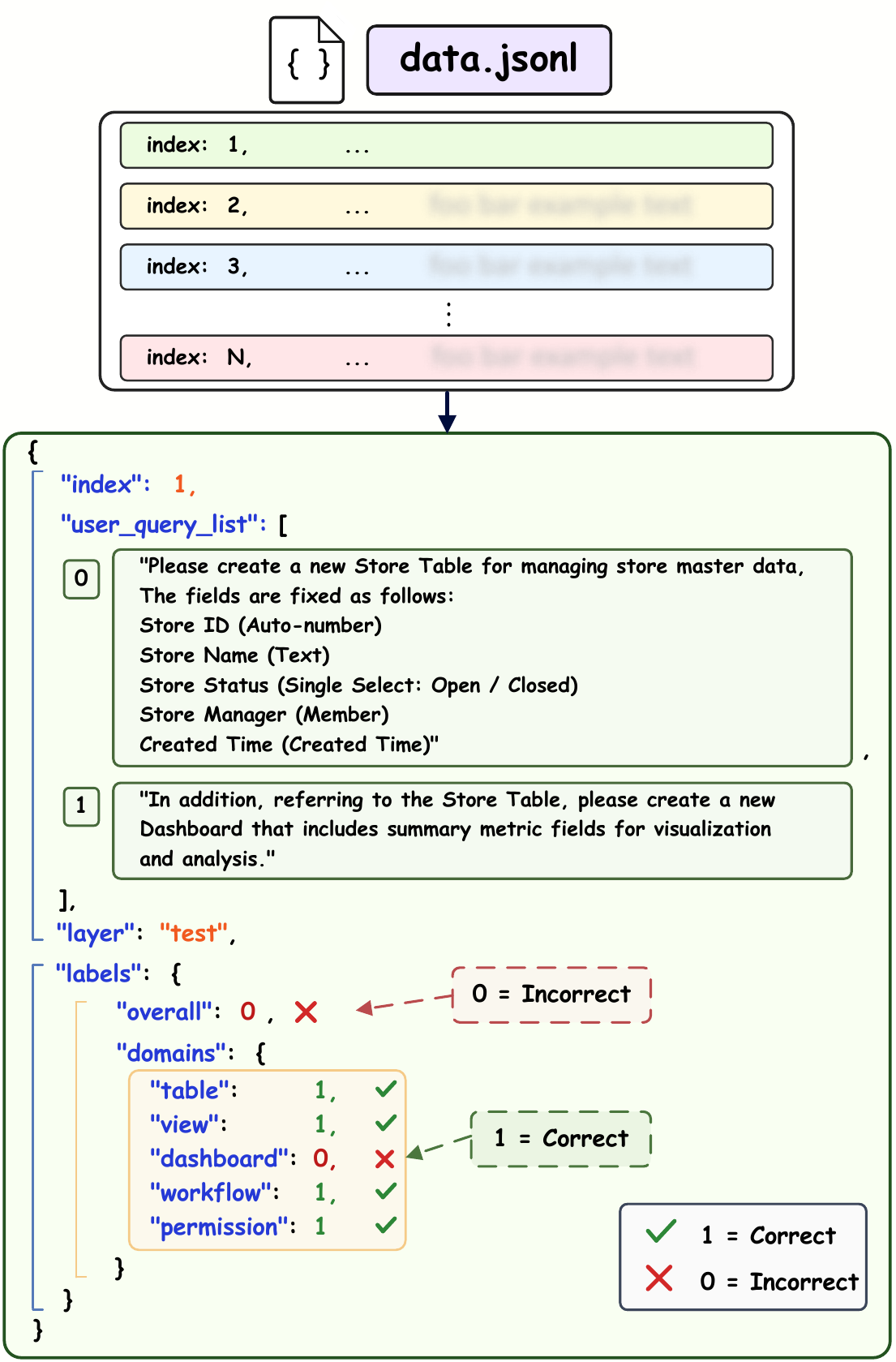}
    \caption{Example JSONL record for a multi-turn spreadsheet-construction trajectory.}
    \label{fig:jsonl}
\end{figure}

\begin{table}[t]
    \centering
    \small
    \setlength{\tabcolsep}{4pt}
    \begin{tabular}{llll}
        \toprule
        \textbf{Stage} & \textbf{Data} & \textbf{Count} & \textbf{Purpose} \\
        \midrule
        SFT & QA & 31,997 ex. / 652 cases & Tool use \\
        SFT & Traj. & 10,059 ex. / 205 cases & Repair \\
        Eval. & Tasks & 595 cases / 1,750 turns & Test \\
        \bottomrule
    \end{tabular}
    \caption{Dataset composition from real usage scenarios used in SheetCraft.}
    \label{tab:dataset}
\end{table}

\begin{table}[t]
    \centering
    \small
    \setlength{\tabcolsep}{4pt}
    \begin{tabular}{p{1.9cm}p{1.1cm}p{3.8cm}}
        \toprule
        \textbf{Metric} & \textbf{Scope} & \textbf{Definition} \\
        \midrule
        All-turns C/E & Case & All turns in a case compile/execute. \\
        Turn-level C/E & Turn & Individual turns compile/execute. \\
        Avg. Dur. & Case & Mean wall-clock time per case. \\
        Case-level Pass@$K$ & Case & Case succeeds within $K$ attempts. \\
        Turn-level Pass@$K$ & Turn & Turn succeeds within $K$ attempts. \\
        Domain pass & Domain & Artifact judge score by component. \\
        \bottomrule
    \end{tabular}
    \caption{Evaluation metrics used in the main comparison and ablation study. C/E denotes compile/execute success.}
    \label{tab:metrics}
\end{table}

\subsection{Dataset}
Our data are derived from real usage scenarios in which users construct and refine spreadsheet artifacts through natural language instructions. Table 1 summarizes the data used in our experiments, where ex. denotes training examples. The SFT data contain two complementary sources: QA examples for learning the spreadsheet operation interface and trajectory examples for learning multi-turn construction and repair. Both sources are distilled from GPT-5.5-Pro and GLM-5.1. We split train and test cases at the case level, stratified by the number of user turns, so that the held-out set preserves the distribution of single-turn and multi-turn tasks. The held-out evaluation set contains 595 real-scenario cases with 1,750 user turns and is used only for inference-time evaluation.

Figure 3 shows the JSONL format of a multi-turn trajectory from the real-scenario data. Each line corresponds to one case and stores task metadata, ordered user turns, before and after spreadsheet artifacts, and artifact-level labels.

\subsection{Evaluation Metrics}
For each user turn, the model generates TypeScript actions that are checked by a compile gate and then executed against a cloned spreadsheet base. We report each main metric as a compile/execute pair, denoted as C/E in Tables 3 and 5. Because a case may contain multiple user turns, we evaluate success at both case and turn granularity. Case-level metrics require all turns in the case to succeed, while turn-level metrics count each user turn independently. We also report Pass@$K$ to measure how often a model solves a turn or case within the first $K$ attempts under the same repair budget.

Table 2 defines the metrics used in our result tables. The main comparison and ablation study use the same reporting format, so model-level results and training-source variants can be compared directly. The domain pass table is complementary. It applies the artifact judge to the final spreadsheet state and shows which spreadsheet components are solved or missed.

\subsection{Settings}
We compare SheetCraft against five strong general-purpose code-agent baselines: GPT-5.5, GPT-5.5-Pro, GLM-5.1, DeepSeek-V4-Pro, and Kimi-K2.5. All systems are evaluated on the same 595 held-out spreadsheet-construction cases sampled from real usage scenarios. To keep the comparison controlled, we fix the system prompt, serialized spreadsheet context, operation documentation, TypeScript interface, retry budget, and compile/execute/judge pipeline across systems. Each case starts from a freshly cloned spreadsheet base, and the model must generate executable TypeScript code that modifies the base through the provided operations.

We choose Qwen3.6-27B as the base model and obtain SheetCraft-27B through supervised fine-tuning. Before training, QA examples and trajectory examples derived from real usage scenarios are converted into the same chat format used at inference time. QA examples contain operation documentation and an operation-level question, while trajectory examples contain the user turn, current spreadsheet schema, relevant operation documentation, prior feedback, and the target assistant program. We compute the language-modeling loss only on assistant answer or code tokens, and treat user requests, schemas, tool documentation, and feedback as conditioning context. This processing makes the two data sources compatible while preserving the distinction between operation knowledge and feedback-conditioned repair.

Training proceeds in two stages on $8\times$ A100 (80GB) GPUs. The first stage trains a QA adapter on operation-level examples so that the model learns operation signatures, argument constraints, and common dependencies between spreadsheet components. The second stage continues SFT on multi-turn trajectory data. Both stages use parameter-efficient LoRA rather than full fine-tuning. In the final trajectory stage, we use LoRA with bfloat16 precision, LoRA rank 16, LoRA alpha 32, dropout 0, and attention projection targets $q, k, v, o$. The maximum sequence length is 23,552 tokens. We train for two epochs with per-device batch size 1, gradient accumulation 3, and effective global batch size 24. The optimizer is AdamW with learning rate $1 \times 10^{-4}$, cosine decay, warmup ratio 0.03, $\beta_1 = 0.9$, $\beta_2 = 0.95$, weight decay 0.1, and maximum gradient norm 1.0. For this stage, we use a five-shard length-grouped strategy: the 10,059 training trajectories are split into shards of 2,012 / 2,012 / 2,011 / 2,012 / 2,012 examples, preserving inter-shard order while sorting examples by tokenized sequence length within each shard. This reduces straggler effects during distributed long-context training. At inference time, we serve the 27B model with vLLM \citep{kwon2023vllm} and keep the prompt template, operation definitions, context budget, maximum generation length, and retry budget fixed. We use temperature 0.2, top-$p$ 1.0, and a maximum context length of 128K tokens for each model call. Each case shares a total budget of at most 20 compile/execute repair attempts (\texttt{maxRunAttempts = 20} in Figure~\ref{fig:pipeline}), and we report Pass@$K$ at $K \in \{1, 3, 5, 10\}$ under this budget; failed model requests are retried twice at the transport layer.

\begin{table*}[t]
    \centering
    \scriptsize
    \setlength{\tabcolsep}{3.2pt}
    \begin{tabular}{lccccccccccc}
        \toprule
        \textbf{Model} & \textbf{All-turns} & \textbf{Turn-level} & \textbf{Avg.} & \multicolumn{4}{c}{\textbf{Case-level Pass@K}} & \multicolumn{4}{c}{\textbf{Turn-level Pass@K}} \\
        \cmidrule(lr){5-8} \cmidrule(lr){9-12}
         & \textbf{C/E} & \textbf{C/E} & \textbf{Dur.} & \textbf{@1} & \textbf{@3} & \textbf{@5} & \textbf{@10} & \textbf{@1} & \textbf{@3} & \textbf{@5} & \textbf{@10} \\
        \midrule
        GPT-5.5 & 94.9/91.6 & 98.0/96.6 & 321s & 31.1/26.9 & 74.8/55.5 & 92.4/80.7 & 94.9/91.6 & 56.9/47.4 & 89.1/78.3 & 96.9/91.4 & 98.0/96.6 \\
        GPT-5.5-Pro & \best{97.5}/\second{94.8} & \best{99.1}/\second{98.2} & 352s & \best{34.5/30.3} & \best{78.2/59.7} & \second{94.1/83.2} & \best{97.5}/\second{94.8} & \best{60.0/51.1} & \best{92.3/82.6} & 97.4/\second{93.1} & \best{99.1}/\second{98.2} \\
        GLM-5.1 & 95.8/93.3 & 98.6/97.6 & 296s & \second{32.8/28.6} & 75.6/57.1 & \second{94.1}/82.4 & 95.8/93.3 & \second{58.3/49.7} & 90.9/80.9 & \second{97.7}/92.6 & 98.6/97.6 \\
        DeepSeek-V4-Pro & 92.4/89.1 & 97.1/95.4 & 302s & 28.6/24.4 & 70.6/52.1 & 89.1/77.3 & 92.4/89.1 & 53.1/44.3 & 86.3/75.1 & 95.4/89.1 & 97.1/95.4 \\
        Kimi-K2.5 & 90.8/86.6 & 96.3/94.0 & 294s & 27.7/22.7 & 68.1/49.6 & 87.4/74.8 & 90.8/86.6 & 51.4/42.9 & 84.6/72.9 & 94.0/87.4 & 96.3/94.0 \\
        \textbf{SheetCraft-27B} & \second{96.6}/\best{95.0} & \second{98.9}/\best{98.3} & 310s & 30.3/27.7 & \second{77.3/58.8} & \best{95.0/84.0} & \second{96.6}/\best{95.0} & 57.4/48.0 & \second{91.7/82.0} & \best{98.0/93.7} & \second{98.9}/\best{98.3} \\
        \bottomrule
    \end{tabular}
    \caption{Multi-turn compile and execute results on the 595-case held-out evaluation set. C/E denotes compile/execute pass rates in percent. All-turns success is computed at case level, turn-level success is computed over 1,750 user turns, and each Pass@K cell reports compile/execute success under the corresponding inference budget. Best results are shown in \best{blue bold}, and second-best results are shown in \second{red bold}.}
    \label{tab:main_results}
\end{table*}

\begin{table}[t]
    \centering
    \scriptsize
    \setlength{\tabcolsep}{2.2pt}
    \begin{tabular}{lccccc}
        \toprule
        \textbf{Model} & \textbf{Overall} & \textbf{Table} & \textbf{Dashboard} & \textbf{Workflow} & \textbf{Permission} \\
        \midrule
        GPT-5.5 & 56.1 & 92.3 & 73.8 & 68.7 & 80.5 \\
        GPT-5.5-Pro & 62.9 & \textbf{96.5} & 78.8 & \textbf{72.9} & 84.7 \\
        GLM-5.1 & 52.8 & 89.7 & 71.3 & 67.1 & 78.8 \\
        DeepSeek-V4-Pro & 49.4 & 87.2 & 68.7 & 63.7 & 75.5 \\
        Kimi-K2.5 & 47.7 & 84.7 & 66.2 & 62.0 & 73.8 \\
        \textbf{SheetCraft-27B} & \textbf{63.1} & 95.8 & \textbf{79.0} & 72.4 & \textbf{85.0} \\
        \bottomrule
    \end{tabular}
    \caption{Artifact-level judge pass rates by spreadsheet domain. Values are percentages. Best results in each column are shown in \textbf{bold}.}
    \label{tab:domain}
\end{table}

\subsection{Evaluation Results}
Table 3 summarizes the main multi-turn results on 595 held-out cases from real usage scenarios with 1,750 user turns. SheetCraft reaches 96.6/95.0 all-turns compile/execute success and 98.9/98.3 turn-level compile/execute success, corresponding to 575/595 all-turns compile success, 565/595 all-turns execute success, 1,730/1,750 turn-level compile success, and 1,720/1,750 turn-level execute success. These results show that the model is reliable at the individual-turn level, while the stricter case-level metric still reveals the difficulty of maintaining correctness across a multi-turn construction process.

The Pass@$K$ columns further highlight the role of repair. At low budget, strong baseline models such as GPT-5.5-Pro and GLM-5.1 show stronger first-attempt performance. With additional attempts, SheetCraft becomes competitive and obtains the best high-budget execution results in several columns, including case-level Pass@5 and turn-level Pass@5/Pass@10. This pattern suggests that supervised trajectory learning mainly improves feedback-conditioned correction. The model is not only learning to generate spreadsheet code, but also learning how to revise it from compiler and execution diagnostics.

Table 4 reports artifact-level judge scores by spreadsheet domain. SheetCraft obtains the strongest overall artifact pass rate and leads on dashboard and permission components. Its table and workflow scores remain close to GPT-5.5-Pro, which is expected because these components often depend on directly specified structure and long-range state. The strongest relative gains appear in dashboard and permission components, where correctness depends on operation ordering, identifier reuse, and repair after execution feedback.

\begin{figure*}[t]
    \centering
    \includegraphics[width=\textwidth]{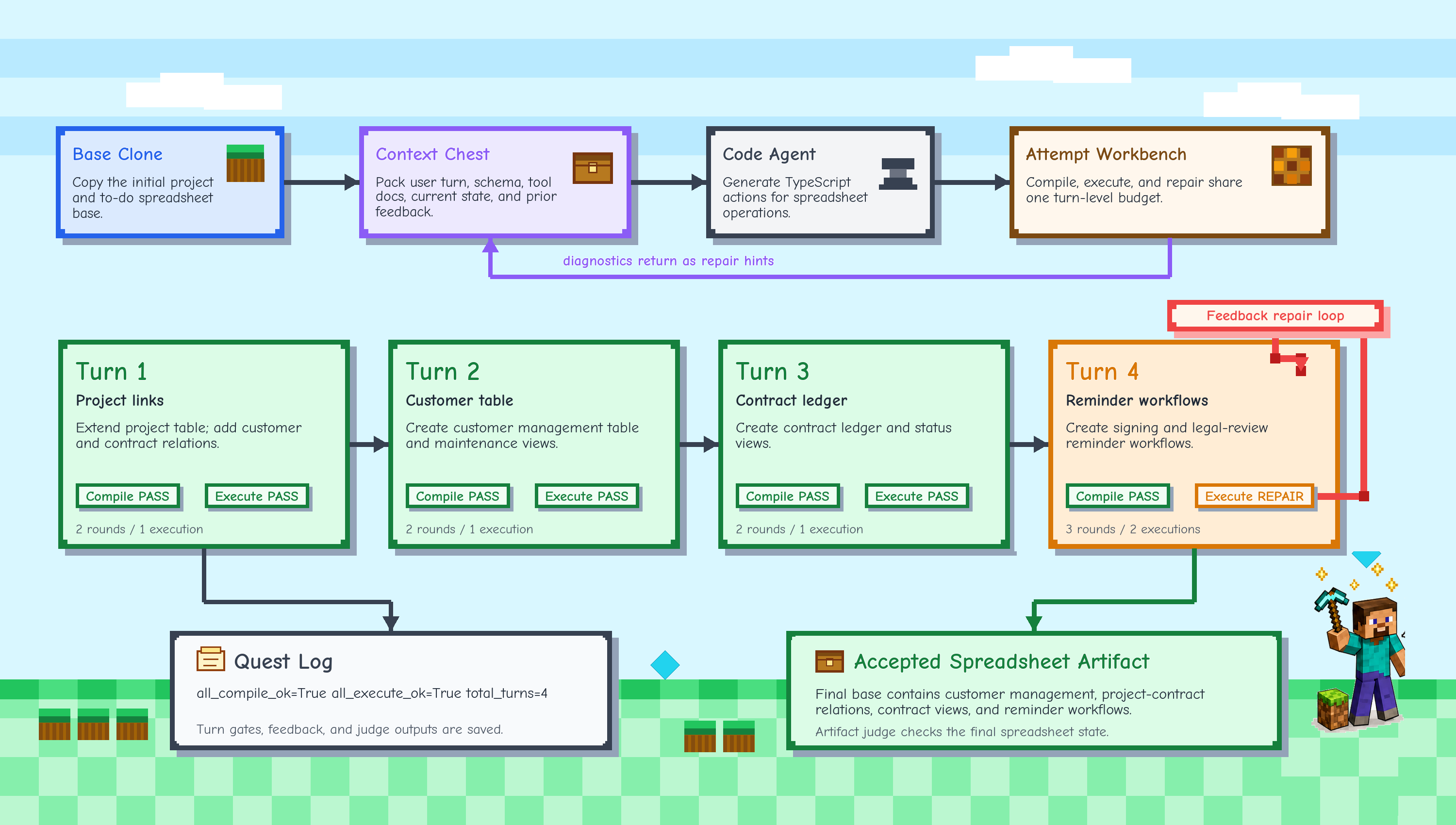}
    \caption{Visualization of a representative four-turn inference trajectory. The selected case successfully passes compilation and execution across all turns; the final workflow turn requires one repair attempt before producing the accepted spreadsheet artifact.}
    \label{fig:case_study}
\end{figure*}

\begin{table*}[t]
    \centering
    \scriptsize
    \setlength{\tabcolsep}{3.2pt}
    \begin{tabular}{lccccccccccc}
        \toprule
        \textbf{Model} & \textbf{All-turns} & \textbf{Turn-level} & \textbf{Avg.} & \multicolumn{4}{c}{\textbf{Case-level Pass@K}} & \multicolumn{4}{c}{\textbf{Turn-level Pass@K}} \\
        \cmidrule(lr){5-8} \cmidrule(lr){9-12}
         & \textbf{C/E} & \textbf{C/E} & \textbf{Dur.} & \textbf{@1} & \textbf{@3} & \textbf{@5} & \textbf{@10} & \textbf{@1} & \textbf{@3} & \textbf{@5} & \textbf{@10} \\
        \midrule
        QA adapter only & 78.2/72.3 & 89.4/84.6 & 248s & 21.0/17.6 & 50.4/36.1 & 69.7/55.5 & 78.2/72.3 & 41.1/33.4 & 68.9/56.3 & 82.6/74.9 & 89.4/84.6 \\
        + GPT-5.5-Pro traj. & \second{94.1/91.6} & \second{97.7/96.0} & 318s & \second{29.4/25.2} & \second{73.1/55.5} & \second{91.6/80.7} & \second{94.1/91.6} & \second{55.4/46.6} & \second{88.9/78.0} & \second{96.6/91.1} & \second{97.7/96.0} \\
        + GLM-5.1 traj. & 91.6/88.2 & 96.3/94.6 & 302s & 27.7/23.5 & 69.7/52.1 & 88.2/76.5 & 91.6/88.2 & 52.9/44.0 & 86.9/75.1 & 94.9/88.9 & 96.3/94.6 \\
        + both teachers & \best{96.6/95.0} & \best{98.9/98.3} & 310s & \best{30.3/27.7} & \best{77.3/58.8} & \best{95.0/84.0} & \best{96.6/95.0} & \best{57.4/48.0} & \best{91.7/82.0} & \best{98.0/93.7} & \best{98.9/98.3} \\
        \bottomrule
    \end{tabular}
    \caption{Ablation results for distillation sources on the 595-case held-out evaluation set. C/E denotes compile/execute pass rates in percent. Each Pass@K cell reports compile/execute success under the corresponding inference budget. Best results are shown in \best{blue bold}, and second-best results are shown in \second{red bold}.}
    \label{tab:ablation}
\end{table*}

\subsection{Ablation Study}
Table 5 isolates the effect of training data sources. The QA-adapter-only variant performs substantially worse than all trajectory-trained variants, especially on case-level Pass@$K$. This indicates that operation-level knowledge is necessary but not sufficient for multi-turn spreadsheet construction. The agent also needs supervision that shows how to react to compile and execution feedback.

Adding trajectory data from either teacher improves both compile and execute success, with GPT-5.5-Pro trajectories giving stronger results than GLM-5.1 trajectories across most columns. Combining both teachers yields the best performance throughout the table. The improvement is most visible under larger Pass@$K$ budgets, which suggests that mixed trajectory data strengthens iterative repair behavior rather than only improving the first generated program.

\begin{table}[t]
    \centering
    \scriptsize
    \setlength{\tabcolsep}{4pt}
    \begin{tabular}{lrrrrr}
        \toprule
        \textbf{Model} & \textbf{In} & \textbf{Out} & \textbf{Mix} & \textbf{E@10} & \textbf{E/\$} \\
        \midrule
        GPT-5.5 & 5.00 & 30.00 & 17.50 & 91.6 & 5.23 \\
        GPT-5.5-Pro & 30.00 & 180.00 & 105.00 & 94.8 & 0.90 \\
        GLM-5.1 & 1.40 & 4.40 & 2.90 & 93.3 & 32.17 \\
        DeepSeek-V4-Pro & 0.435 & 0.87 & 0.6525 & 89.1 & 136.55 \\
        Kimi-K2.5 & 0.60 & 3.00 & 1.80 & 86.6 & 48.11 \\
        \bottomrule
    \end{tabular}
    \caption{API price comparison for public baseline models. In, Out, and Mix are USD per one million tokens; Mix is the 1:1 input-output cost proxy. E@10 is case-level execute Pass@10, and E/\$ is E@10 divided by Mix. DeepSeek-V4-Pro uses cache-miss input pricing.}
    \label{tab:pricing}
\end{table}

\subsection{API Cost Comparison}
The public baselines also differ substantially in inference cost. To make this trade-off visible, Table 6 reports the listed API prices of the baseline models and aligns them with the case-level execute Pass@10 metric from Table 3. Prices are taken from the official OpenAI release note, Z.AI pricing page, DeepSeek pricing page, and Kimi API platform. We use a simple 1:1 input-output cost proxy,
\[
c_{\mathrm{mix}} = \frac{c_{\mathrm{in}} + c_{\mathrm{out}}}{2},
\]
and report $\text{E@10}/c_{\text{mix}}$ as a cost-normalized reading of execution success.

\section{Case Study and Visualization}
Figure 4 expands one held-out case into its inference-time execution flow. The case contains four user turns and starts from an existing project and to-do spreadsheet base. The agent first extends the project table with customer and contract links, then adds customer management views, contract ledger views, and finally contract-signing and legal-review reminder workflows. The case highlights a core challenge of spreadsheet construction: later turns must reuse and extend identifiers, schemas, and views created in earlier turns.

The case study also shows how execution feedback enters the repair loop. Each user turn is converted into a TypeScript program under the same context-construction and attempt-budget mechanism. Compilation and execution results are logged at the turn level, and runtime feedback is used for targeted repair when needed. The final outcome depends on both the accumulated turn-level gates and the spreadsheet artifact produced on the cloned base.

\section{Conclusion}
We presented SheetCraft, an execution-guided framework for training and evaluating code agents that construct persistent spreadsheet artifacts from natural language instructions. SheetCraft combines tool-use QA supervision with multi-turn trajectory supervision derived from real usage scenarios, and deploys the resulting agent through a budgeted repair loop grounded in compilation diagnostics, sandbox execution, and artifact-level judging. Experiments on a 595-case held-out benchmark from real usage scenarios show that robust spreadsheet construction depends on spreadsheet-specific operation knowledge, feedback-driven repair, and evaluation of the final artifact state. These results suggest that execution-grounded trajectory learning is a promising direction for building practical agents for structured productivity workflows.

\section*{Limitations}
SheetCraft is evaluated on real usage scenarios for spreadsheet construction, but the benchmark is still limited in scale and domain coverage. The 595 held-out cases emphasize business-oriented artifacts and may not cover all spreadsheet ecosystems, specialized workflows, or multilingual instructions. Because the framework relies on executable feedback from cloned spreadsheet bases, results can also be affected by external service availability, authentication, cloning behavior, and transient execution failures. Artifact-level judging is more faithful than code-only evaluation, but it still cannot fully capture layout quality, interaction preferences, or ambiguous business intent. Finally, long multi-turn histories remain a practical bottleneck because they increase latency and may exceed the inference context budget, motivating future work on state abstraction, history summarization, and stronger long-horizon recovery policies.

\bibliography{custom}

\end{document}